\documentclass[dvips,sissa]{prl}

\begin{document}

\newcommand{\noi}{\noindent}
\newcommand{\ie}{\textit{i.e.,\xspace}}
\newcommand{\eg}{\textit{e.g.,\xspace}}
\newcommand{\etc}{\textit{etc.\xspace}}
\newcommand{\cf}{\textit{cf.\xspace}}
\newcommand{\etal}{\textit{et al.\xspace}}
\newcommand{\half}{\frac{1}{2}}
\newcommand{\third}{\frac{1}{3}}
\newcommand{\fourth}{\frac{1}{4}}
\newcommand{\sss}{\scriptscriptstyle}

\renewcommand{\Im}{\text{Im}}
\renewcommand{\Re}{\text{Re}}
\renewcommand{\vec}[1]{\ensuremath{\mathbf{#1}}}

\title{
Coupling to optical phonons in the one-dimensional t-J model: \\
Effects on superconducting fluctuations and phase separation
}

\author{ 
R. Fehrenbacher$^*$ \\
\itshape  Materials Science Division, 
Argonne National Laboratory, Argonne, Illinois 60439
}

\newaddress{$^*$ Present address: Max-Planck-Institut f\"ur
Festk\"orperforschung, Heisenbergstr. 1, D-70569 Stuttgart, Germany.}

\newcommand{\sj}{{\hspace{-1.3pt} j}}

\newcommand{\ybco}{YBa$_2$Cu$_3$O$_{7-y}$\xspace}
\newcommand{\htc}{high-$T_c$\xspace}
\newcommand{\dw}{$d_{x^2 - y^2}$\xspace}
\newcommand{\cuo}{CuO$_2$\xspace}
\newcommand{\tjm}{$t$-$J$ model\xspace}

\newcommand{\cs}[1]{\ensuremath{c_{#1\sigma}}}
\newcommand{\csd}[1]{\ensuremath{c_{#1\sigma}^\dagger}}
\newcommand{\csdd}[1]{\ensuremath{c_{#1\downarrow}^\dagger}}
\newcommand{\csdu}[1]{\ensuremath{c_{#1\uparrow}^\dagger}}
\newcommand{\calp}[1]{\ensuremath{\mathcal{P}_{#1}}}
\newcommand{\sumijs}{\ensuremath{\sum_{\langle i,j \rangle, \sigma}}}  
\newcommand{\sumij}{\ensuremath{\sum_{\langle i,j \rangle}}}
\newcommand{\udispl}{\ensuremath{|0\rangle_i^{\phantom{\dagger}}}}
\newcommand{\displ}{\ensuremath{|\widetilde 0\rangle_i^{\phantom{\dagger}}}}
\newcommand{\kro}{\ensuremath{K_\rho}\xspace}
\newcommand{\ebc}{\ensuremath{E_{\hspace{-0.5pt}B}^{\hspace{0.8pt}c}}\xspace}

\maketitle

\begin{abstract}
\noi 
The one-dimensional (1D) $t$-$J$ Holstein model is studied by exact
diagonalization of finite rings using a variational approximation for the
phonon states. Due to renormalization effects induced by the phonons, for
intermediate electron-phonon coupling, the phase separation (PS) boundary, and
with it the region of dominating superconducting fluctuations is shifted
substantially to smaller values of $J/t$ as compared to the pure
\tjm. Superconducting correlations are weakened through charge density wave
interactions mediated by the phonons. Possible consequences for the \htc oxides
are discussed. \\

\noi PACS numbers: 74.20.Mn, 74.25.Kc, 71.27.+a, 71.38.+i
\end{abstract}

\begin{multicols}{2}

The role of the electron-phonon (e-ph) coupling for the mechanism leading to
superconductivity (SC) in the \htc cuprates is still controversial.  The
importance of electronic correlations suggests itself from the fact that the
\htc is achieved upon doping a Mott insulator. Hence, many theoretical
approaches neglect the phonons altogether and search for a purely electronic
mechanism. An obvious shortcoming of such theories is the fact that they would
never lead to an isotope effect. 

While the experimental situation concerning the magnitude of the isotope
effect and the relevant ions producing it is still not quite settled, it
appears to be widely accepted by now that especially for underdoped samples,
$T_c$ is rather susceptible especially to the mass of the oxygen ions
\cite{zech-etal-94}. Near optimal doping, however, only a small isotope
effect is observed. In addition, anomalous frequency shifts of phonon modes as
well as other lattice anomalies have been observed at or slightly above $T_c$
\cite{egami-billinge-94}. These findings strongly suggest that the electronic
and phonon degrees of freedom should be treated on an equal footing.

The simplest model to describe the {\em electronic} low-energy properties of
doped \cuo sheets is the \tjm \cite{zhang-rice-88}. Due to elimination of
doubly occupied sites, it describes \emph{strongly correlated} electrons for
all parameters. This explains the extreme difficulties to obtain reliable
results by analytical as well as numerical methods. An exception is 1D, where
an exact solution exists at $J/t = 2$
\cite{bares-blatter-90,*sutherland-75,*schlottmann-87}. Numerically, it was
possible to extract the full phase diagram with good accuracy
\cite{ogata-etal-91}. In brief, there are three distinct regions: For $J/t
\lesssim 2.2$, the physics is dominated by charge density wave (CDW)
fluctuations, followed by a small regime ($2.2 \lesssim J/t \lesssim 3.2$)
where SC fluctuations are strongest, and finally, at large $J/t$, phase
separation sets in. In addition, at low electron density, a spin gap opens in
the regime of dominating SC fluctuations.

In this article, we study the influence of the e-ph coupling on the phase
diagram of the 1D $t$-$J$ model. In particular, it is important to know how
this affects the SC fluctuations. The e-ph coupling is incorporated by a
Holstein term, \ie a linear coupling of the local electronic charge to the
ionic displacement. Together with the free phonon part, we arrive at the
$t$-$J$ Holstein model. Its phase diagram is calculated by exact
diagonalization. The phonons are treated by a variational approximation which
{\em preserves} their dynamics. The main results are: (i) Increasing the e-ph
coupling shifts the regime of dominating SC fluctuations to smaller values of
$J/t$. (ii) The PS boundary acquires a non-trivial density dependence due to an
enhancement of CDW correlations by the e-ph coupling, and (iii) the spatial
extent of SC correlations is shortened.

The formation of a Zhang-Rice (ZR) singlet (representing a Cu$^{\text{III}}$
oxidation state) by holes doped in the insulating half-filled state of the \cuo
sheets is the central idea underlying the derivation of the $t$-$J$ model
from the more complete three-band model \cite{zhang-rice-88,hybertsen-etal-90}.
We have demonstrated previously in the context of CuO$_3$ chains
\cite{fehrenbacher-94a}, that the total energy $E_s$ of this singlet is
sensitive to displacements of the planar oxygen towards the Cu atoms (the
so-called breathing mode) through the distance dependence of the hopping matrix
elements $t_{pd}$ and $t_{pp}$. The dependence of $E_s$ on the displacement is
particularly strong for an in-phase motion, since the doped hole
is predominantly in a state delocalized with {\em equal} probability on the
four oxygens surrounding a Cu. A rough estimate of the coupling
(as in \cite{fehrenbacher-94a}) gives a value of $g \equiv
dE_s/dQ \approx$ 10eV/\AA ($Q$ the displacement).

Experimentally, $g$ can be estimated from the difference $\delta Q$ of the Cu-O
bond lengths of compounds with pure four-fold oxygen coordinated
Cu$^{\text{II}}$ (\eg Nd$_2$CuO$_4$, $d_{\text{Cu-O}} \approx 1.97$\AA
\cite{lightfoot-etal-90}), versus Cu$^{\text{III}}$ (\eg KCuO$_2$,
$d_{\text{Cu-O}} \approx 1.83$\AA \cite{brese-etal-89}). From this, $g = \delta
Q K \approx 10 - 15$eV/\AA\xspace, in good agreement with the theoretical
estimate. $K = \Omega^2 M \approx 100$eV/\AA$^2$ is the spring constant
obtained from the measured phonon frequencies \cite{pintschovius-etal-91}($M =
4M_{\text{Ox}}$, $M_{\text{Ox}}$ the oxygen atomic mass).

The simplest model which incorporates the dependence $E_s (Q)$ is the 
$t$-$J$ Holstein model \cite{fehrenbacher-94a}
\begin{equation}
\begin{split}
H & = - t \sumijs \calp{i} \csd{i} \cs{\sj} \calp{\sj} + 
        J \sumij \left( \vec{S}_i \cdot \vec{S}_{\sj} - 
                        \fourth n_i n_{\sj} \right) \\
& + \quad 
        \hbar \Omega \sum_i a_i^\dagger a_i + 
        \Lambda \sum_{i} \left( 1 - n_i \right) 
                                \left( a_i + a_i^\dagger\right).
\end{split}
\label{t-J-Hol-ham}
\end{equation}
where $\langle i, j \rangle$ denotes nearest neighbors (nn), the \cs{i},
\csd{i} annihilate (create) an electron with spin $\sigma$ at site $i$, 
$n_i = \sum_\sigma n_{i\sigma}$ is the number, and $\vec{S}_i$ the
spin operator. The projection operators $\calp{i} = 1 - n_{i\uparrow}
n_{i\downarrow}$ eliminate configurations with doubly occupied sites, and the
$a_i, a_i^\dagger$ are usual bosonic phonon annihilation (creation)
operators. For simplicity, the phonon spectrum is assumed to be dispersionless,
characterized by a single frequency $\Omega$. The e-ph coupling constant is
defined as $\Lambda = (\hbar g^2 / 2 M \Omega )^{1/2}$.

The total Hamiltonian has four dimensionless parameters: the spin exchange
$J/t$, the phonon frequency $\hbar\Omega/t$, the e-ph coupling strength
$\Lambda / t$, and the electron density $0 \leq n \leq 1$. The energy scale is
fixed by the value of $t$.  Here, we choose $\hbar\Omega/t = 0.2$,
realistic for the cuprates ($t \approx 0.4$eV
\cite{hybertsen-etal-90}, and $\Omega \approx 60-80$meV for the breathing modes
\cite{pintschovius-etal-91}). We measure the e-ph coupling in
terms of the polaron binding energy, $E_B \equiv g^2/(2K) =
\Lambda^2/(\hbar\Omega )$ ($-E_B$ is the energy of a localized polaron),
instead of $\Lambda / t$.  Using the above values for $g$, $\Omega$, we obtain
$E_B \approx 0.1-0.4$eV, \ie a sizeable $E_B/t \approx 0.25-1$. This
corresponds to a rather small value $\lambda \equiv E_B / W \approx 0.1 - 0.2$
($W$, the bandwidth in 2D).

The model is studied using the Lanczos method on finite rings of length $L$. To
minimize finite-size effects, we chose closed shell boundary conditions (BC),
which prevent spurious degeneracies. For single-band 1D models, this
corresponds to periodic BC if the number of electrons $n_e = 4m + 2$, and
antiperiodic BC if $n_e = 4m$ ($m$ a positive integer). Then the ground state
is a spin singlet.

The infinite-dimensional phonon Hilbert space (for each mode) requires some
further approximation even for arbitrarily small systems. Previous approaches
to similar models \cite{zhong-schuettler-92,roeder-fehske-buettner-93} treated
the phonons in the {\em static} limit, \ie classically. The ground state within
this limit is obtained by minimizing the electronic energy with respect to the
lattice displacement (Born-Oppenheimer). However, neglecting the phonon
dynamics can lead to a rather severe {\em overestimate} of the tendency towards
small polaron formation, \ie self-trapping of the carriers, especially for the
large optical phonon frequencies found in the cuprates.

An unbiased approximation {\em preserving the phonon dynamics} results from
a truncation of the Hilbert space to include $n_p$ phonon
states per site. We have shown \cite{fehrenbacher-95b}, that for the
simplest case, $n_p = 2$, there is a {\em natural and unique} choice for the
two states at site $i$, $|\phi_1 \rangle_i, |\phi_2 \rangle_i$ defined in terms
of the phonon vacuum \udispl and the {\em
coherent} (displaced oscillator) state $\displ = e^{-\eta^2 / 2} e^{\eta
a_i^\dagger} \udispl$ ($\eta^2 = E_B/(\hbar \Omega)$)
\begin{equation}
|\phi_1 \rangle_i = \frac{1}{\sqrt{1 - e^{-\eta^2}}} \left( \udispl - 
e^{-\eta^2 / 2} \displ \right); \; |\phi_2 \rangle_i = \displ \; .
\end{equation}
The {\em non-perturbative} character of this ansatz allows for asymptotically
exact ground state properties in the weak {\em and} strong coupling limit. For
intermediate coupling, the approximation was checked by including additional
phonon states (see Ref. \citen{fehrenbacher-94a}). Apart from an upward
renormalization of the coupling constant, we found no changes of ground state
properties (electronic properties of the system with additional phonon states
at $E_B$ are identical within an error $< 0.1\%$ to the properties of the
system with only $|\phi_1 \rangle_i, |\phi_2 \rangle_i$ at $E_B^\prime$, with
$0.55 \lesssim E_B^\prime/E_B < 1$ depending on $E_B$). More details will be
published elsewhere. Here, we shall restrict ourselves to $n_p = 2$ using
$|\phi_1 \rangle_i, |\phi_2 \rangle_i$.

The phase diagram of the model is determined by calculating (i) the
compressibility $\kappa$, and (ii) the critical exponent \kro which determines
the decay of correlation functions in 1D models belonging to either the
Tomonaga-Luttinger liquid (TLL) \cite{solyom-79,*haldane-81,schulz-90} or the
Luther-Emery liquid (LEL) \cite{luther-emery-74} universality class. Both
systems are characterized by gapless charge excitations. The spin excitations
are gaped in a LEL, while they remain gapless in a TLL.

The transition to a phase separated state is signaled by a divergent
$\kappa$, which we calculate from
\begin{equation}
\kappa = \frac{L}{n_e^2}\frac{4}{E_0 (L, n_e + 2) + E_0 (L, n_e - 2) - 2
E_0 (L, n_e)} \; ,
\end{equation}
$E_0 (L, n_e)$ the ground state energy of a ring with $n_e$ electrons and $L$
sites. \kro is given by $\kro = (\pi D n^2 \kappa/4)^{1/2}$ ($n = n_e/L$, the
electron density) \cite{schulz-90}, with the Drude weight
\cite{shastry-sutherland-90}
\begin{equation}
D = \pi L \left. \frac{\partial^2 E_0 (\Phi)}{\partial \Phi^2} \right|_{\Phi =
\Phi_0} \; .
\end{equation}
Here, $E_0(\Phi)$ is the ground state energy for BC with a phase factor
$e^{i\Phi}$, and $\Phi_0 = 0,\pi$ for periodic and antiperiodic BC
respectively. A value of $\kro > 1$ indicates that SC correlations dominate the
long-range fluctuations for both a TLL, and a LEL, while $\kro < 1$ is
characteristic for dominating CDW correlations.

To check the consistency of the assumed TLL scaling, we calculated the central
charge $c$ from the finite size corrections of the ground state energy as
predicted by conformal field theory, $E(L)/L = \epsilon_\infty - \pi c (v_c +
v_s)/(6L^2)$ \cite{frahm-korepin-90,*schlottmann-87}. Here $v_c, v_s$ are the
charge and spin velocity which we extract from the dynamical charge and spin
structure factor as in Ref. \citen{tohyama-horsch-maekawa-95}. Using the
results for $L = 8, 12$, we found a maximum deviation of 0.07 from the expected
$c = 1$ for the range of parameters studied. Given the finite-size errors in
the velocities, this confirms the validity of TLL scaling.

In Fig. \ref{fig:phase-diag}, we show the calculated phase diagrams for $E_B =
0, 0.25, 0.5, 0.75$, $L = 10, 12, 16$.  Including the phonons, we were
restricted to $L \leq 12$, but reliable results can already be obtained for $L
= 10$.  The main observations are: (i) With increasing e-ph coupling, the PS
boundary, as determined from the condition $1/\kappa = 0$, is shifted
considerably to lower $J/t$. An independent criterion for PS is the long
wavelength divergence of the charge structure factor $S_{\text{ch}}(k) = 1/L
\sum_{j,l}^L e^{ik(j-l)} \langle n_j n_l \rangle$ which gives an error estimate
$< 5\%$ for the phase boundary.  A similar shift without a significant
reduction in area is observed for the region of dominating SC fluctuations
occurring below the onset of PS. For $E_B = 0.75$, it extends down to values as
low as $J/t \approx 1.0$, compared to $J/t \approx 2.2$ in the pure
\tjm. Qualitatively, this shift can be explained by an {\em enhancement of the
effective mass} of the charge carriers due to the e-ph
coupling, \ie a downward renormalization of $t \mapsto \tilde t \ll t$, which
results in an {\em enhanced effective} $J/{\tilde t}$. Note, that $J$ is not
expected to be appreciably renormalized, since the phonons couple to the
charge. Also, the direct dependence of $J$ on the displacements of the
breathing mode is weak, since out of the two Cu-O bonds between a nn Cu pair,
they strengthen one, but weaken the other one.


\begin{center}
\vspace*{-2.5mm}
\includegraphics{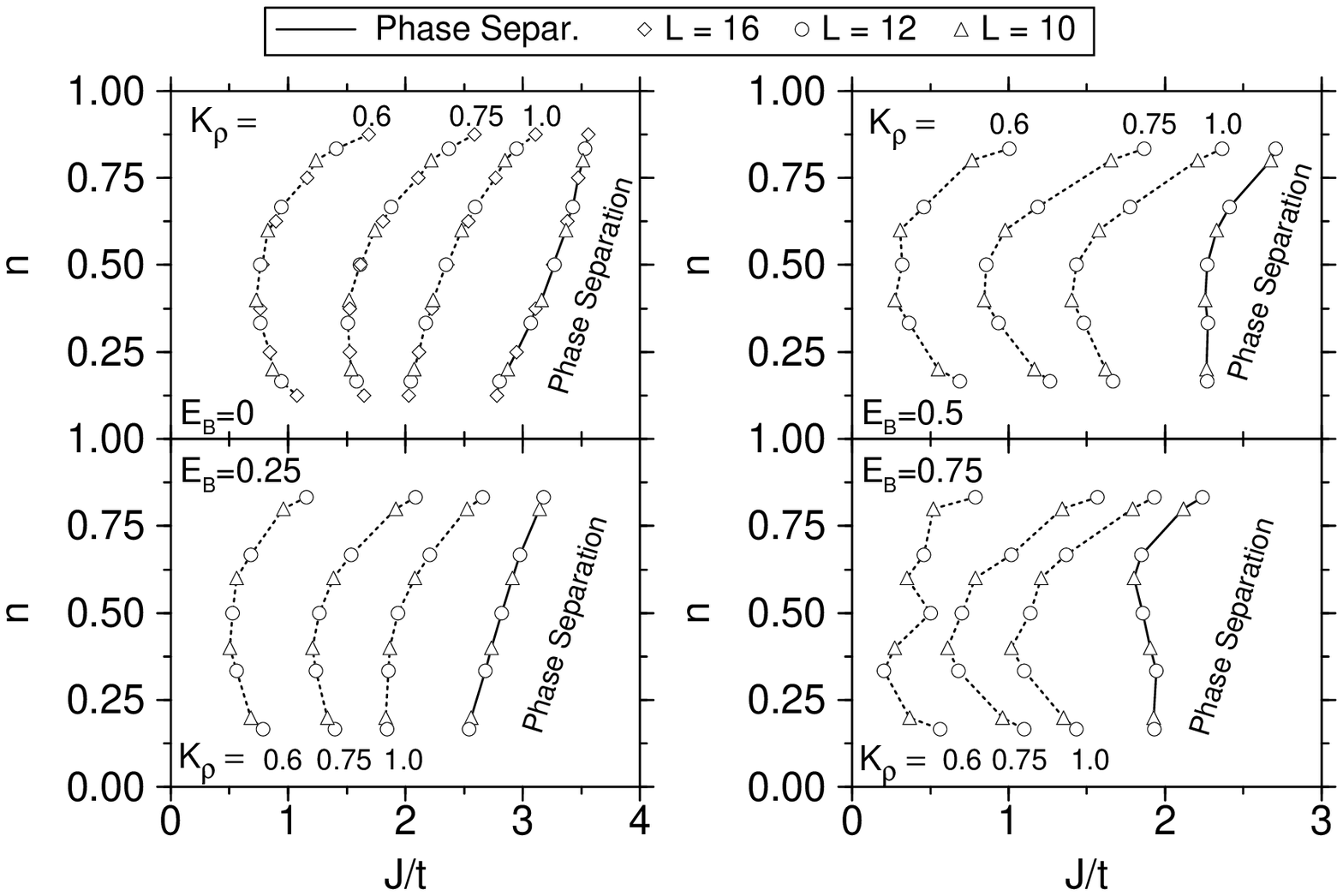}
\vspace{-6mm}
\begin{mycaption}
The phase diagram of the 1D $t$-$J$ Holstein model for $\Omega/t = 0.2$ and
various values of the e-ph coupling $E_B$.
\label{fig:phase-diag}
\end{mycaption}
\vspace{-3mm}
\end{center}
\noi

(ii) Especially at large $E_B$, the PS line and the contours of
constant $\kro$ acquire a {\em strong density dependence}. In particular, a dip
in the $\kro = 0.6$ contour develops for $E_B = 0.75$ at quarter filling. This
is due to enhanced CDW correlations from the e-ph
coupling. In fact, at $E_B = 0.75$, there is {\em
long-range} CDW order for $J/t \lesssim 0.4$, as indicated by a
saturation in the staggered density-density correlation function 
\begin{equation} 
D_{\text{ch}} (j) = \frac{1}{L} \sum_i (-1)^i \left( \langle n_i n_{i+j} 
\rangle - \langle n_i \rangle \langle n_{i+j} \rangle \right) \; ,
\end{equation}
\ie $\lim_{j \to \infty} D_{\text{ch}} (j) \neq 0$.
Note, that for the case $J/t = 0$ (equivalent to spinless fermions),
it is fairly well established by analytical and numerical results
\cite{hirsch-fradkin-83,fehrenbacher-95b} that the system undergoes a CDW
transition at $n = 0.5$ above a {\em finite} critical coupling strength $\ebc
(\Omega)$.  In the static limit, $\ebc = 0$, demonstrating the importance of
quantum fluctuations. Previously \cite{fehrenbacher-95b}, we estimated $\ebc/t
\approx 0.53$ for $\Omega/t = 0.2, n_p = 2$. Note however, that this value is
definitely too low because of the renormalization of the coupling constant by
additional phonon states as explained above. An analysis
of the charge structure factor $S_{\text{ch}}(k) = 1/L \sum_{j,l}^L e^{ik(j-l)}
\langle n_j n_l \rangle$ strongly indicates that the appearance of a
$4k_F$ CDW is not unique to density $n = 0.5$ (which is obvious at the
mean-field level). However, we found that for $n \neq 0.5$, the weakened
lattice commensurability leads to an enhanced $\ebc/t$, as estimated from
the values at which $S_{\text{ch}}(4k_F)$ appears to diverge as a function of
$L$. We conclude that the e-ph coupling in this model mediates
an effectively {\em repulsive} interaction between the electrons, and hence
{\it does not} support SC.

\begin{center}
\includegraphics[width=6.0cm]{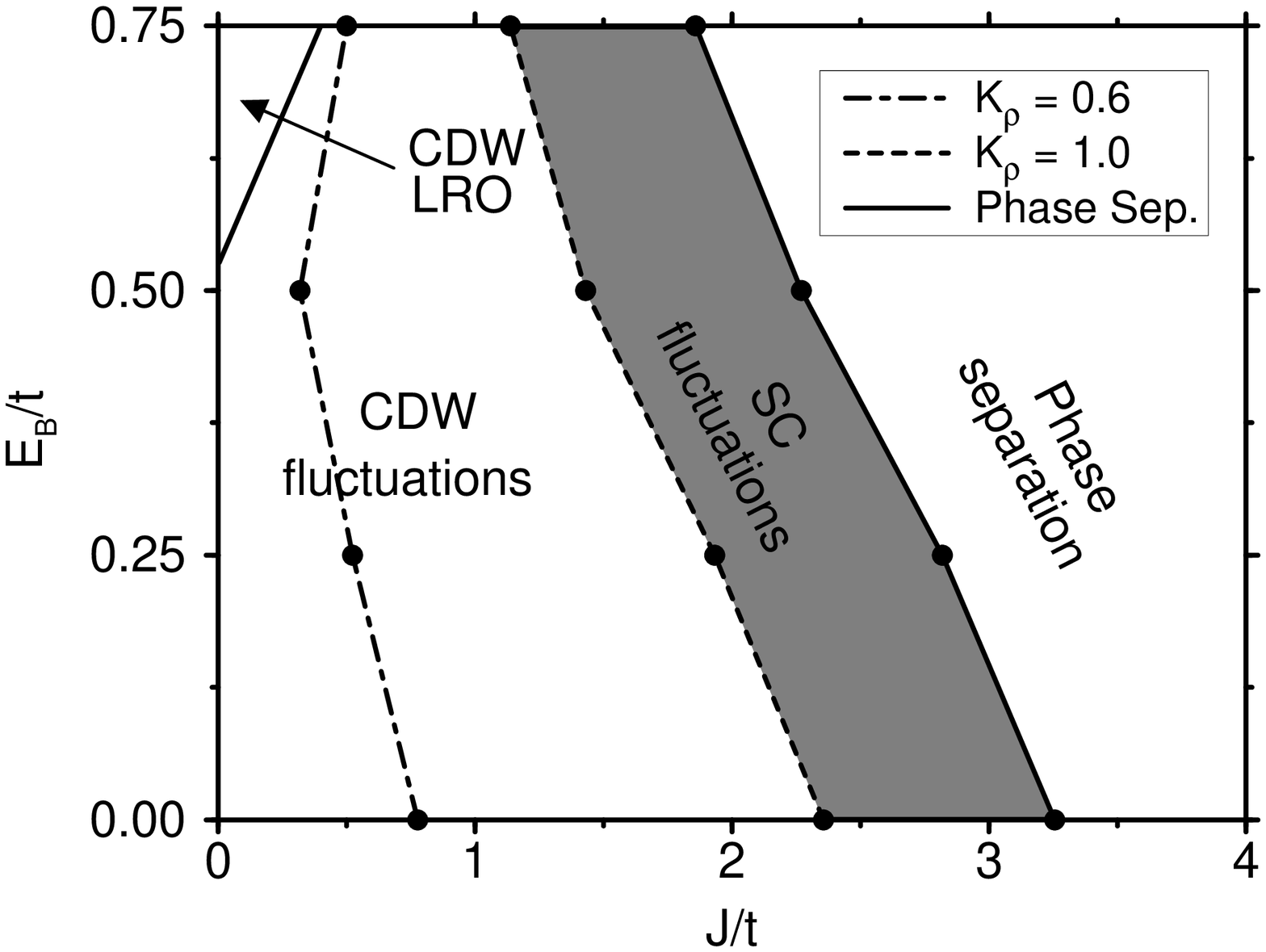}
\vspace*{-3mm}
\begin{mycaption}
The phase diagram of the 1D $t$-$J$ Holstein model in the $E_B, J$ plane at
quarter filling for $\Omega/t = 0.2$, and $L = 12$.
\label{fig:J-pb-phase-diag}
\end{mycaption}
\vspace*{-3mm}
\end{center}


Fig. \ref{fig:J-pb-phase-diag} shows the phase diagram in the $E_B, J$ plane at
quarter filling. This plot clearly demonstrates that the regime of SC
fluctuations is shifted almost {\em rigidly} to smaller values of $J/t$ as the
e-ph coupling is increased. Note, the small region of CDW order at small $J/t$,
large $E_B/t$. Within a strong coupling perturbation expansion for the case of
spinless fermions, it was shown \cite{hirsch-fradkin-83}, that apart from the
well known exponential renormalization of the hopping, $t \to t
e^{-E_B/\Omega}$, the Holstein coupling also leads to an {\em effective nn
repulsion}, which dominates in the limit $E_B \to \infty$. The competition of
this effective phonon-mediated repulsion with the attractive spin exchange term
is responsible for the corresponding strong competition between CDW and SC
correlations in the $t$-$J$ Holstein model. For $E_B > \ebc (J = 0)$, a phase
transition from a CDW to a metal takes place as $J$ is increased from 0.

Next, we analyze the effect of the e-ph coupling on the superconducting
correlations directly, by looking at (i) the singlet SC structure factor
defined as
\begin{equation}
S_{\text{pair}} (k) = \frac{1}{L} \sum_{j,l}^L e^{ik(j-l)} \langle 
P_{\sj}^\dagger P_l \rangle \; ,
\end{equation}
where $P_j^\dagger = (\csdu{\sj} \csdd{\sj +1} - \csdd{\sj} \csdu{\sj +1} )/
\sqrt{2}$ creates a nn singlet pair, and (ii) the pair-pair
correlation function in real space 
$D_{\text{pair}} (i) = \langle P_i^\dagger 
P_0 \rangle$. It is important to study $D_{\text{pair}} (i)$ separately to
distinguish short and long-range correlations. 

In Fig. \ref{fig:pair-corell}(a), we plot $S_{\text{pair}} (k = 0)$ at quarter
filling as a function of $J/t$, $E_B$.  We observe an overall {\em suppression}
of $S_{\text{pair}} (k = 0)$, as $E_B$ increases. This can be traced back to
the repulsive CDW correlations introduced by the phonons. In addition, the peak
at the onset of PS is increasingly smeared out as $E_B$
grows. Looking at the correlations in real space,
Fig. \ref{fig:pair-corell}(b), we note that this suppression is coming
primarily from a reduction of {\em long-range} correlations. This
underlines our conclusion that the e-ph coupling is {\em detrimental} to SC in
this model.

\begin{center}
\vspace*{-2mm}
\includegraphics{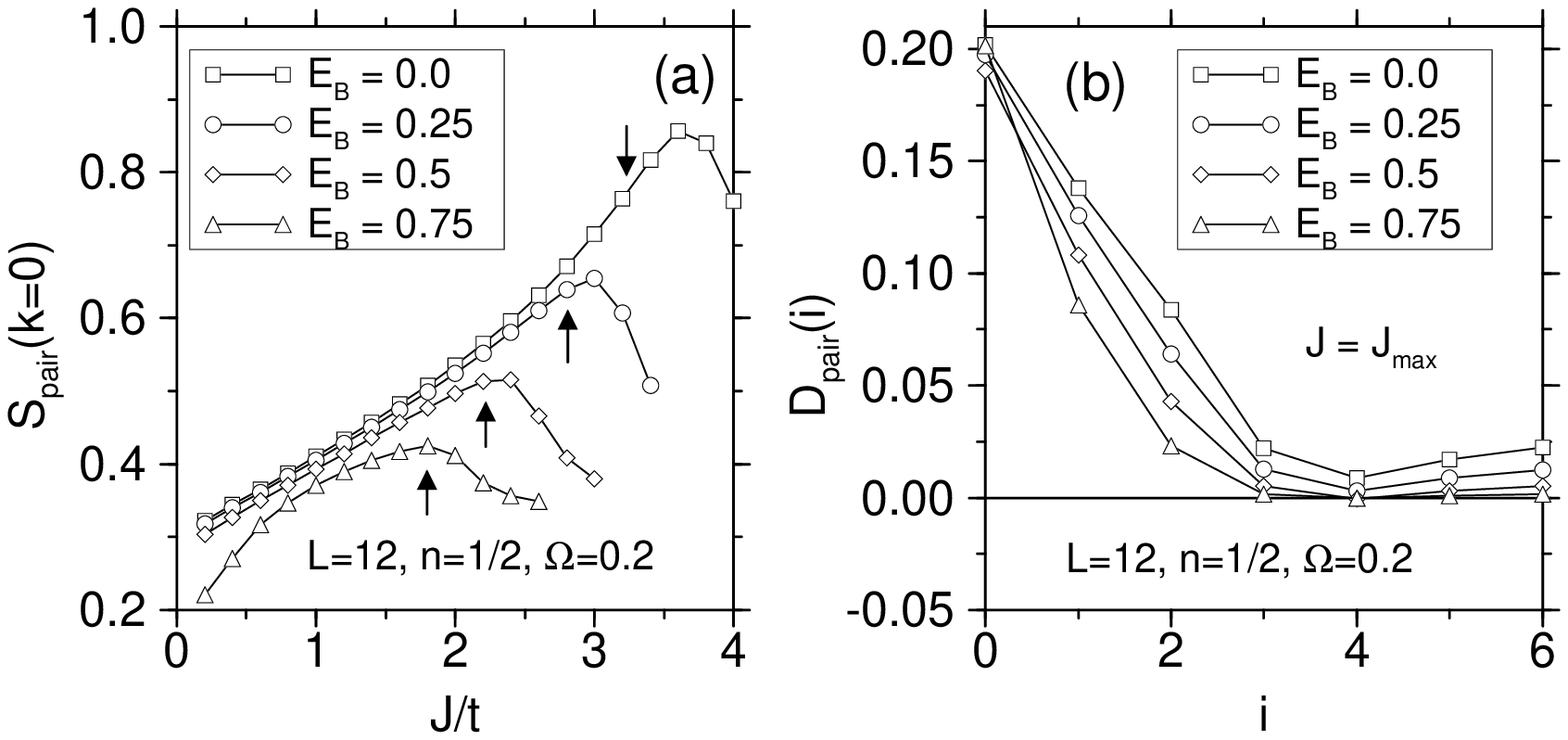}
\vspace*{-4mm}
\begin{mycaption}
(a) The singlet SC structure factor $S_{\text{pair}} (k = 0)$ for various
    $E_B$. The arrows mark the occurrence of PS. (b) The
    pair-pair correlation function $D_{\text{pair}} (i)$ for various $E_B$
    evaluated at the value $J = J_{\text{max}} (E_B)$, where $S_{\text{pair}}
    (k = 0, J, E_B)$ has a maximum.
\label{fig:pair-corell}
\end{mycaption}
\vspace*{-2mm}
\end{center}


Another question concerns the existence of a spin gap $\Delta_s$, defined as
the energy difference between the singlet ground state and the lowest triplet
excitation for $L \to \infty$. It was shown \cite{troyer-etal-93} that the
addition of a nn repulsion to the \tjm can lead to a large $\Delta_s$ in the
region of dominating SC fluctuations. Since the e-ph coupling also induces an
effective short-range repulsion, one might expect an analogous appearance of a
spin gap. Due to the restriction $L \leq 12$, an accurate finite-size scaling
for $\Delta_s$ is not possible. Nevertheless, extrapolating $\Delta_s(1/L)$ for
$L = 8, 12$ at $n = 1/4$ to $1/L = 0$, the intercept is negative or very small
for all $E_B$. However, at $E_B = 0.75$, we {\em consistently} observe a small
positive intercept for several $J$ in the region of dominating SC fluctuations,
possibly indicating a small spin gap. For the $t$-$J$-$V$ model studied in
\cite{troyer-etal-93}, we found that extrapolating results from $L = 8, 12$
calculations also produces a negative intercept as long as $\Delta_s/t \lesssim
0.1$. This is due to the fact that the scaling function is exponential at small
$1/L$ if a spin gap is present \cite{troyer-etal-93}. Assuming the same scaling
of $\Delta_s$ with $1/L$ for the $t$-$J$-$V$ and the present model, we can give
an upper bound $\Delta_s/t \lesssim 0.1$ for all $E_B$ studied.

In the 2D \tjm, {\em coherent} charge motion is already renormalized by
spin-exchange scattering resulting in an effective bandwidth $\sim J$
\cite{dagotto-94}. This should make the e-ph coupling more effective, \ie
probably a smaller coupling (as in 1D) is needed to see appreciable
effects. Substantial evidence for a SC ground state of \dw symmetry in the 2D
\tjm has been reported \cite{dagotto-riera-93,*ohta-etal-94,heeb-rice-94}.  
The most reliable estimate for the SC phase boundary 
predicts a critical value $J_c/t \approx 0.5$ above which SC appears in the
interesting regime of hole doping ($\delta = 1 - n = 0.15-0.25$)
\cite{heeb-rice-94}. Our results suggest that the additional mass
enhancement by the e-ph coupling could push down the phase boundary to lower
$J$, so that SC might arise in the regime of $J/t \approx 1/4 - 1/3$, relevant
for the cuprates \cite{hybertsen-etal-90}. Concerning the isotope effect, our
results are ambiguous. A smaller phonon frequency increases the ratio
$E_B/\Omega$, \ie {\em strengthens} the effect of the e-ph coupling. However,
this has two {\em competing} effects on SC: (i), the effective ratio $J/\tilde
t$ increases through a further mass enhancement, but (ii), the induced
repulsive CDW correlations also grow. The identification of the dominating
effect requires further calculations in 2D.

The author would like to thank M. R. Norman and T. M. Rice for useful comments,
and a careful reading of the manuscript. We acknowledge financial support by
the NSF (NSF-DMR-91-20000) through the Science and
Technology Center for Superconductivity. Numerical calculations have been
performed on the Convex C38, and HP 735 workstations at CSSC Manno,
Switzerland, and on the Cray2 at NERSC, Lawrence Livermore
Lab., CA.

\end{multicols}


\begin{mcbibliography}{10}

\bibitem{zech-etal-94}
D.~Zech et~al.,
\newblock Nature {\bf 371}, 681 (1994)\relax
\relax
\bibitem{egami-billinge-94}
T.~Egami and S.~J.~L. Billinge,
\newblock Prog. Mat. Sci. {\bf 38}, 359 (1994)\relax
\relax
\bibitem{zhang-rice-88}
F.~C. Zhang and T.~M. Rice,
\newblock Phys. Rev. B {\bf 37}, 3759 (1988)\relax
\relax
\bibitem{bares-blatter-90}
P.~A. Bares and G.~Blatter,
\newblock Phys. Rev. Lett. {\bf 64}, 2567 (1990)\relax
\relax
\bibitem{sutherland-75}
B.~Sutherland,
\newblock Phys. Rev. B {\bf 12}, 3795 (1975)\relax
\relax
\bibitem{schlottmann-87}
P.~Schlottmann,
\newblock Phys. Rev. B {\bf 36}, 5177 (1987)\relax
\relax
\bibitem{ogata-etal-91}
M.~Ogata et~al.,
\newblock Phys. Rev. Lett. {\bf 66}, 2388 (1991)\relax
\relax
\bibitem{hybertsen-etal-90}
M.~S. Hybertsen et~al.,
\newblock Phys. Rev. B {\bf 41}, 11068 (1990)\relax
\relax
\bibitem{fehrenbacher-94a}
R.~Fehrenbacher,
\newblock Phys. Rev. B {\bf 49}, 12230 (1994)\relax
\relax
\bibitem{lightfoot-etal-90}
P.~Lightfoot et~al.,
\newblock Physica C {\bf 168}, 627 (1990)\relax
\relax
\bibitem{brese-etal-89}
N.~E. Brese et~al.,
\newblock J. Sol. St. Chem. {\bf 83}, 1 (1989)\relax
\relax
\bibitem{pintschovius-etal-91}
L.~Pintschovius et~al.,
\newblock Physica C {\bf 185-189}, 156 (1991)\relax
\relax
\bibitem{zhong-schuettler-92}
J.~Zhong and H.-B. Sch{\"u}ttler,
\newblock Phys. Rev. Lett. {\bf 69}, 1600 (1992)\relax
\relax
\bibitem{roeder-fehske-buettner-93}
H.~R{\"o}der, H.~Fehske, and H.~B{\"u}ttner,
\newblock Phys. Rev. B {\bf 47}, 12420 (1993)\relax
\relax
\bibitem{fehrenbacher-95b}
R.~Fehrenbacher,
\newblock preprint\relax
\relax
\bibitem{solyom-79}
J.~S{\'o}lyom,
\newblock Adv. Phys. {\bf 28}, 201 (1979)\relax
\relax
\bibitem{haldane-81}
F.~D.~M. Haldane,
\newblock J. Phys. C {\bf 14}, 2585 (1981)\relax
\relax
\bibitem{schulz-90}
H.~J. Schulz,
\newblock Phys. Rev. Lett. {\bf 64}, 2831 (1990)\relax
\relax
\bibitem{luther-emery-74}
A.~Luther and V.~J. Emery,
\newblock Phys. Rev. Lett. {\bf 33}, 589 (1974)\relax
\relax
\bibitem{shastry-sutherland-90}
B.~S. Shastry and B.~Sutherland,
\newblock Phys. Rev. Lett. {\bf 65}, 243 (1990)\relax
\relax
\bibitem{frahm-korepin-90}
H.~Frahm and V.~E. Korepin,
\newblock Phys. Rev. B {\bf 42}, 10553 (1990)\relax
\relax
\bibitem{tohyama-horsch-maekawa-95}
T.~Tohyama, P.~Horsch, and S.~Maekawa,
\newblock Phys. Rev. Lett. {\bf 74}, 980 (1995)\relax
\relax
\bibitem{hirsch-fradkin-83}
J.~E. Hirsch and E.~Fradkin,
\newblock Phys. Rev. B {\bf 27}, 4302 (1983)\relax
\relax
\bibitem{troyer-etal-93}
M.~Troyer et~al.,
\newblock Phys. Rev. B {\bf 48}, 4002 (1993)\relax
\relax
\bibitem{dagotto-94}
E.~Dagotto,
\newblock Rev. Mod. Phys. {\bf 66}, 763 (1994)\relax
\relax
\bibitem{dagotto-riera-93}
E.~Dagotto and J.~Riera,
\newblock Phys. Rev. Lett. {\bf 70}, 682 (1993)\relax
\relax
\bibitem{ohta-etal-94}
Y.~Ohta et~al.,
\newblock Phys. Rev. Lett. {\bf 73}, 324 (1994)\relax
\relax
\bibitem{heeb-rice-94}
E.~S. Heeb and T.~M. Rice,
\newblock Europhys. Lett. {\bf 27}, 673 (1994)\relax
\relax
\end{mcbibliography}
\end{document}